\documentclass[12pt]{article}
\usepackage{epsfig}
\usepackage{amsfonts}
\usepackage{amsmath}
\usepackage{amssymb}
\usepackage{cite}
\usepackage{rotating}
\begin{document}
\sffamily
\title
{Extracting the resonance parameters from experimental data on scattering of
charged particles}
\author{P. Vaandrager, S.A. Rakityansky$^*$\\[3mm]
\parbox{11cm}{
{\small Dept. of Physics, University of Pretoria, Pretoria, South Africa\\
* e-mail: rakitsa@up.ac.za
}
}}
\maketitle
\begin{abstract}
\noindent
A new parametrization of the multi-channel $S$-matrix is used to fit
scattering data and then to locate the resonances as its poles. The $S$-matrix
is written in terms of the corresponding ``in'' and ``out'' Jost matrices which
are expanded in the Taylor series of the collision energy $E$ around an
appropriately chosen energy $E_0$. In order to do this, the Jost matrices are
written in a semi-analytic form where all the factors (involving the channel
momenta and Sommerfeld parameters) responsible for their ``bad behaviour'' (i.e.
responsible for the multi-valuedness of the Jost matrices and for branching of
the Riemann surface of the energy) are given explicitly. The remaining unknown
factors in the Jost matrices are analytic and single-valued functions of the
variable $E$ and are defined on a simple energy plane. The expansion is done for
these analytic functions and the expansion coefficients are used as the fitting
parameters. The method is tested on a two-channel model, using a set of
artificially generated data points with typical error bars and a typical random
noise in the positions of the points.
\end{abstract}
\bigskip
Published in:\\
{
International Journal of Modern Physics E,
Vol. 25, No. 2 (2016) 1650014 (12 pages)\\
DOI: 10.1142/S0218301316500142
}
\section{Introduction}
The main parameters that characterize any quantum resonance, are the
collision energy $E_r$, at which this state can be excited, and the width
$\Gamma$ that determines the lifetime of the state. For a multi-channel system
the total width is the sum of the partial widths,
$\Gamma=\Gamma_1+\Gamma_2+\cdots$, where $\Gamma_n/\Gamma$ gives the relative
probability of decaying into the $n$-th channel. There are many different
methods for determining these parameters from a set of scattering data (several
of them are described in Refs.~\cite{Kukulin, NSTAR}). These
methods form two big groups based on two principally different approaches.\\

Within one approach, the parameters $E_r$, $\Gamma$, $\Gamma_1$, $\Gamma_2$, etc
are treated as the adjustable variables in a procedure of fitting the available
experimental data. The simplest and most well-known example of such a method is
the Breit-Wigner parametrization of the amplitude~\cite{BW36}. A common feature
of all of the methods belonging to this category, is that the number of
resonances is fixed from the outset. All these methods use some parametric
expression for the amplitude, or for the $S$-matrix, or directly for the cross
section, where the resonance singularities (or the zigzags of the cross section)
are embedded into this parametric expression by hand. These methods only differ
in the method of parametrization and in the derivation of the parametric
expression.\\

Within the second approach, the resonances are considered as the poles of the
$S$-matrix at the complex energies $E_r-i\Gamma/2$ in an appropriate domain of
the Riemann surface of the energy. The $S$-matrix is written in a more general
form with some adjustable parameters that do not necessarily coincide with the
resonance parameters. Usually, it is not known beforehand how many
resonances can be found (if any). After fitting the data at real collision
energies, the analytic expression for the $S$-matrix thus obtained, is examined
at complex energies where the poles (if found) are interpreted as the
resonances. The Pad\'e approximation of the $S$-matrix~\cite{Kukulin, ourPade,
Pade1}  and the Laurent-Pietarinen series expansion of the
amplitude~\cite{Svarc1} can be mentioned as examples.\\

The method we describe here belongs to the second category and is based on the
rigorous semi-analytic expression for the $N$-channel Jost matrix derived in
Ref.~\cite{ourCoulomb}. In that expression, all the factors responsible for the
``bad behaviour'' of the Jost matrix (i.e. factors depending on the Sommerfeld
parameters and the channel momenta responsible for the branching of the Riemann
surface) are given explicitly. The remaining unknown factors are analytic and
single-valued functions of $E$ defined on a simple energy plane. These functions
are expanded in the Taylor series, and the expansion coefficients serve as the
fitting parameters.

\section{Parametrization}
In Ref.~\cite{ourCoulomb}, it was shown that for a non-relativistic reaction of
the type $a+b\to c+d$ involving charged particles, the $N$-channel Jost
matrix has the following general form:
\begin{eqnarray}
\label{multi.matrixelements}
   f^{(\mathrm{in/out})}_{mn}(E) &=&
   \frac{e^{\pi\eta_m/2}\ell_m!}{2\Gamma(\ell_m+1\pm i\eta_m)}
   \left\{
   \frac{C_{\ell_n}(\eta_n)k_n^{\ell_n+1}}
   {C_{\ell_m}(\eta_m)k_m^{\ell_m+1}}{A}_{mn}(E)\ -\right.\\[3mm]
\nonumber
   &-&
   \left.\left[
   \frac{2\eta_mh(\eta_m)}{C_0^2(\eta_m)}\pm i\right]
   C_{\ell_m}(\eta_m)C_{\ell_n}(\eta_n)
   k_m^{\ell_m}k_n^{\ell_n+1}{B}_{mn}(E)\right\}\ ,
\end{eqnarray}
where
\begin{equation}
\label{chmom}
    k_n=\pm\sqrt{\frac{2\mu_n}{\hbar^2}(E-E_n)}\ ,\qquad n=1,2,\dots,N\ ,
\end{equation}
are the channel momenta determined by the differences between the total energy
$E$ and the channel thresholds $E_n$, as well as by the corresponding reduced
masses $\mu_n$; the channel angular momenta and the Sommerfeld parameters are
$\ell_n$ and $\eta_n=\mu_n e^2Z_1Z_2/(k_n\hbar^2)$; the function
\begin{equation}
\label{CoulombC}
   C_\ell(\eta)=\frac{2^\ell e^{-\pi\eta/2}}{\Gamma(2\ell+2)}
   \left|\Gamma(\ell+1\pm i\eta)\right|
\end{equation}
is the Coulomb barrier factor; and
\begin{equation}
\label{h_function}
   h(\eta)=\frac12\left[\psi(i\eta)+
   \psi(-i\eta)\right]-\ln \hat{\eta}\ ,
   \qquad
   \psi(z)=\frac{\Gamma'(z)}{\Gamma(z)}\ ,
   \qquad
   \hat{\eta}=\frac{\mu e^2|Z_1Z_2|}{k\hbar^2}\ .
\end{equation}
It was shown that the remaining unknown matrices $A(E)$ and $B(E)$ in
Eq.~(\ref{multi.matrixelements}) are single-valued and analytic functions of the
energy, defined on a simple energy plane without branching points. All the
complicated topology of the Riemann surface where the Jost functions are
defined, is determined by the coefficients of the matrices $A(E)$ and $B(E)$,
given in Eq.~(\ref{multi.matrixelements}) explicitly.\\

If for a given energy $E$ the Jost matrices (\ref{multi.matrixelements}) are
known, then the corresponding $S$-matrix is just their ``ratio'',
\begin{equation}
\label{S_matrix}
    S(E)=f^{\rm (out)}(E)\left[f^{\rm (in)}(E)\right]^{-1}\ ,
\end{equation}
and the scattering cross section for the channel $n\to m$ can be found as
\begin{equation}
\label{cross_section_sigma_nm}
   \sigma_{mn}(E)=
   \frac{\pi}{k_n^2}(2\ell_n+1)\left|S_{mn}(E)-\delta_{mn}\right|^2\ .
\end{equation}
The resonances are the points
\begin{equation}
\label{Res_energy}
   {\cal E}=E_r-\frac{i}{2}\Gamma\ ,\qquad
   E_r>0\ ,\quad \Gamma>0\ ,
\end{equation}
on the Riemann surface of the energy, where
\begin{equation}
\label{spectral}
    \det f^{\rm (in)}({\cal E})=0
\end{equation}
and therefore where the $S$-matrix has poles.\\

The energy surface has a square-root branching point at every channel threshold
$E_n$. This is because the Jost matrices depend on the energy $E$ via the
channel momenta (\ref{chmom}) and for each of them there are two
possible choices of the sign in front of the square root. The resonance
spectral points are located on the so called non-physical sheet of this Riemann
surface, i.e. such a layer of the surface where all the channel momenta have
negative imaginary parts. In the numerical calculations, the choice of the
sheet is done by an appropriate choice of the signs in front of the square
roots (\ref{chmom}).\\

Since the matrices $A(E)$ and $B(E)$ are analytic, they can be expanded in the
Taylor series around any complex point $E_0$. Near this point, they can
therefore be approximated by the first $M$ terms of these series:
\begin{eqnarray}
\label{A_series}
   A(E) &\approx&
   \sum_{i=0}^M a_i(E_0)(E-E_0)^i\ ,\\[3mm]
\label{B_series}
   B(E) &\approx&
   \sum_{i=0}^M b_i(E_0)(E-E_0)^i\ ,
\end{eqnarray}
where the expansion coefficients $a_i$ and $b_i$ are
($N\times N$)-matrices. These matrices depend only on the choice of the point
$E_0$. After finding them, the Jost matrices (\ref{multi.matrixelements}) can
be used at any complex energy $E$ within a circle around $E_0$ where the
approximations (\ref{A_series}, \ref{B_series}) are satisfactory.\\

We treat the elements of the matrices $a_i$ and $b_i$ as the
adjustable parameters in the procedure of fitting experimental cross section.
After finding the optimal values for them, we look for the roots of
Eq.~(\ref{spectral}) and thus find the resonance parameters $E_r$ and $\Gamma$.
As to the partial widths $\Gamma_n$, they can easily be found following the
procedure described in Ref.~\cite{Qchem}. Indeed, we know their sum
$\Gamma=\Gamma_1+\Gamma_2+\cdots+\Gamma_N$ and we can find their ratios (see
Ref.~\cite{Qchem}):
\begin{equation}
\label{GGratio}
   \frac{\Gamma_m}{\Gamma_n}=\lim_{E\to{\cal E}}
   \left|\frac{S_{mm}(E)}{S_{nn}(E)}\right|\ .
\end{equation}
At a resonance energy, $E={\cal E}$, all the elements of the $S$-matrix are
singular because all of them have the same singular factor
$1/\det f^{\rm (in)}(E)$. However, in the ratio (\ref{GGratio}) this factor
cancels out. Therefore, if we explicitly invert the matrix $f^{\rm (in)}$ and
use it in Eq.~(\ref{S_matrix}) without common factor $1/\det f^{\rm (in)}$,
then we can avoid numerical evaluation of the limit (\ref{GGratio}). In the
simplest case of a two-channel problem, we obtain:
\begin{equation}
\label{GGratio2}
   \frac{\Gamma_1}{\Gamma_2}=\left|
   \frac{f^{\rm (out)}_{11}f^{\rm (in)}_{22}-
         f^{\rm (out)}_{12}f^{\rm (in)}_{21}}
        {f^{\rm (out)}_{22}f^{\rm (in)}_{11}-
         f^{\rm (out)}_{21}f^{\rm (in)}_{12}}\right|_{E=\mathcal{E}}\ ,
    \qquad
    N=2\ .
\end{equation}
It should be noted that such a simple and numerically stable procedure for
calculating the partial widths is only possible when the Jost matrices are
parametrized. If we were parametrizing  the $S$-matrix directly, then the limits
(\ref{GGratio}) would have to be calculated numerically as the ratios of
singular functions.

\section{Fitting}
We assume that there are  sets of experimental data available for at least one
channel $n\to m$ (or perhaps for several channels), i.e. the cross sections
$$
   \sigma_{mn}\left(E^{(mn)}_i\right) \pm \delta^{(mn)}_i\ ,
   \qquad
   i=1,2,\dots,N^{(mn)}
$$
with the corresponding experimental errors (standard deviations)
$\delta^{(mn)}$, measured at the collision energies $E^{(mn)}_i$. The center
$E_0$ of the expansions  (\ref{A_series}, \ref{B_series}) can be chosen
somewhere within the interval covered by these collision energies
(where we expect to find a resonance). The optimal values of the expansion
parameters are found by minimizing the following $\chi^2$ function
\begin{eqnarray}
\label{chi2}
   \chi^2 &=&
   \displaystyle
   \sum_{i=1}^{N^{(mn)}}\left[\frac{\sigma_{mn}(E^{(mn)}_i)-
   \sigma^{\mathrm{fit}}_{mn}(E^{(mn)}_i)}
   {\delta^{(mn)}_i}\right]^2\\[3mm]
\nonumber
   &+&
   \displaystyle
   \sum_{j=1}^{N^{(m'n')}}\left[\frac{\sigma_{m'n'}(E^{(m'n')}_j)-
   \sigma^{\mathrm{fit}}_{m'n'}(E^{(m'n')}_j)}
   {\delta^{(m'n')}_j}\right]^2
   \ +\ \cdots\\[3mm]
\nonumber
   &+&
   \displaystyle
   \sum_{m<n,j}\left|S_{mn}(E^{(mn)}_j)-S_{nm}(E^{(mn)}_j)\right|^2\ ,
\end{eqnarray}
where the fitting cross section $\sigma^{\mathrm{fit}}_{mn}$
depends on the expansion coefficients via Eqs.
(\ref{cross_section_sigma_nm}), (\ref{S_matrix}) and
(\ref{multi.matrixelements}). The last sum in the above $\chi^2$ function makes
the fitted $S$-matrix symmetric in accordance with the detailed balance
theorem (see a more detailed discussion in Ref.~\cite{ourFit}). The
minimization is done using the MINUIT code~\cite{minuit1, minuit2}.\\

The number of the adjustable parameters depends on the number $N$ of the
existing channels (the data do not have to be available for all of them) and on
the number $M$ of the terms in the Taylor series (\ref{A_series},
\ref{B_series}). Generally speaking, the expansion coefficients $a(E_0)$ and
$b(E_0)$ are the $N\times N$ matrices of complex elements. However, as was
shown in Ref.~\cite{ourCoulomb}, they become real matrices if the point $E_0$
is on the real axis. Therefore, if $E_0$ is on the interval covered by
the experimental energies $E^{(mn)}_i$, the the total number of real fitting
parameters is $2(M+1)N^2$.

\section{Example}
In order to demonstrate the efficiency of the proposed method, we choose a model
two-channel problem where the parameters of the resonances can be determined
in an exact way. For this model, we generate artificial data points with a
typical distribution of errors. In addition to the error-bar for each
pseudo-data point, we introduce a random shift (up or down) from the exact
cross section curve, i.e. a typical experimental ``noise''. Using these points,
we extract the resonance parameters and compare them with the corresponding
exact values.\\

Our artificial data points are generated using the following two-channel
potential having a Coulomb tail:
\begin{equation}
\label{pot}
   V(r)= \left(
         \begin{array}{cc}
         -1.0 & -7.5 \\
         -7.5 & 7.5
         \end{array} \right) r^{2} e^{-r}+
         \left(
         \begin{array}{cc}
         1 & 0 \\
         0 & 1   \end{array} \right)
         \frac{1}{r}\ .
\end{equation}
The short-range term in this potential is the same as in the famous Noro-Taylor
model~\cite{norotaylor}. The units in Eq.~(\ref{pot}) are therefore the same,
namely, they are such that the reduced masses for both channels are equal to
one, $\mu_1=\mu_2=1$, with $\hbar c=1$, and both angular momenta are zero,
$\ell_1=\ell_2=0$. The threshold energies for the channels are
$E_1=0$ and $E_2=0.1$.\\
\begin{table}[htbp]
\centering
\begin{tabular}{| c || c | c | c | c |}
\hline
 & $E_{r}$ & $\Gamma$ & $\Gamma_{1}$ & $\Gamma_{2}$ \\ [1ex]
\hline\hline
1 & 6.278042551      & 0.036866729   & 0.006898807     & 0.029967922 \\ [1ex]
\hline
2 & 8.038507867      & 2.563111275   & 0.617710684     & 1.945400591 \\ [1ex]
\hline
3 & 8.861433400      & 7.883809113   & 1.949506410     & 5.934302704 \\ [1ex]
\hline
4 & 9.020824224      & 14.07907263   & 3.591961102     & 10.48711153 \\ [1ex]
\hline
5 & 8.566130944      & 20.75266055   & 5.414178669     & 15.33848188 \\ [1ex]
\hline
6 & 7.548492959      & 27.69926473   & 7.328979882     & 20.37028485 \\ [1ex]
\hline
\end{tabular}
\caption{The resonance energies and widths generated by the potential
(\ref{pot}). }
\label{table.exact_poles}
\end{table}

For a given potential, the matrices $A(E)$ and $B(E)$ in
Eq.~(\ref{multi.matrixelements}) can be found as the solutions of differential
equations derived in Ref.~\cite{ourCoulomb}. This can be done for any complex
energy with the help of the complex rotation of the coordinate described in
Ref.~\cite{Qchem}. In this way we can find the exact Jost matrices and
therefore the exact cross section as well as the exact resonance parameters. The
first six resonances thus located for the potential (\ref{pot}) are listed in
Table~\ref{table.exact_poles}.\\

For each of the elastic channels, $(1\to1)$ and $(2\to2)$, we generated 30
artificial data points in the energy interval $6<E<11$. In order to make them
more realistic, these points were randomly shifted around the corresponding
exact cross section curves, using the Gaussian distribution, i.e. the values
$\sigma_{mn}(E^{(mn)}_i)$ were replaced with
$$
    \sigma_{mn}(E^{(mn)}_i)\ \longrightarrow
    \ \sigma_{mn}(E^{(mn)}_i)G_i\ ,
$$
where $G_i$ were the normally distributed random numbers with the mean value $1$
and the standard deviation $\Delta$. We used three values of $\Delta$, namely,
$0.01$, $0.05$, and $0.10$. This was done to test the stability of the method.\\

The center of expansion was taken as $E_0=8$. In the case of low experimental
noise ($\Delta=0.01$), we used $M=5$, i.e. the first six terms of the series
(\ref{A_series}, \ref{B_series}) were taken into account. For higher noise, the
number of terms in the expansions was smaller, namely, $M=3$. The reason for
such a choice was that with larger $M$ the fitting curve tries to pass  through
almost all the data points and thus does noisy zigzags, which result in a loss
of overall accuracy.\\

\begin{figure}[ht!]
\centerline{\epsfig{file=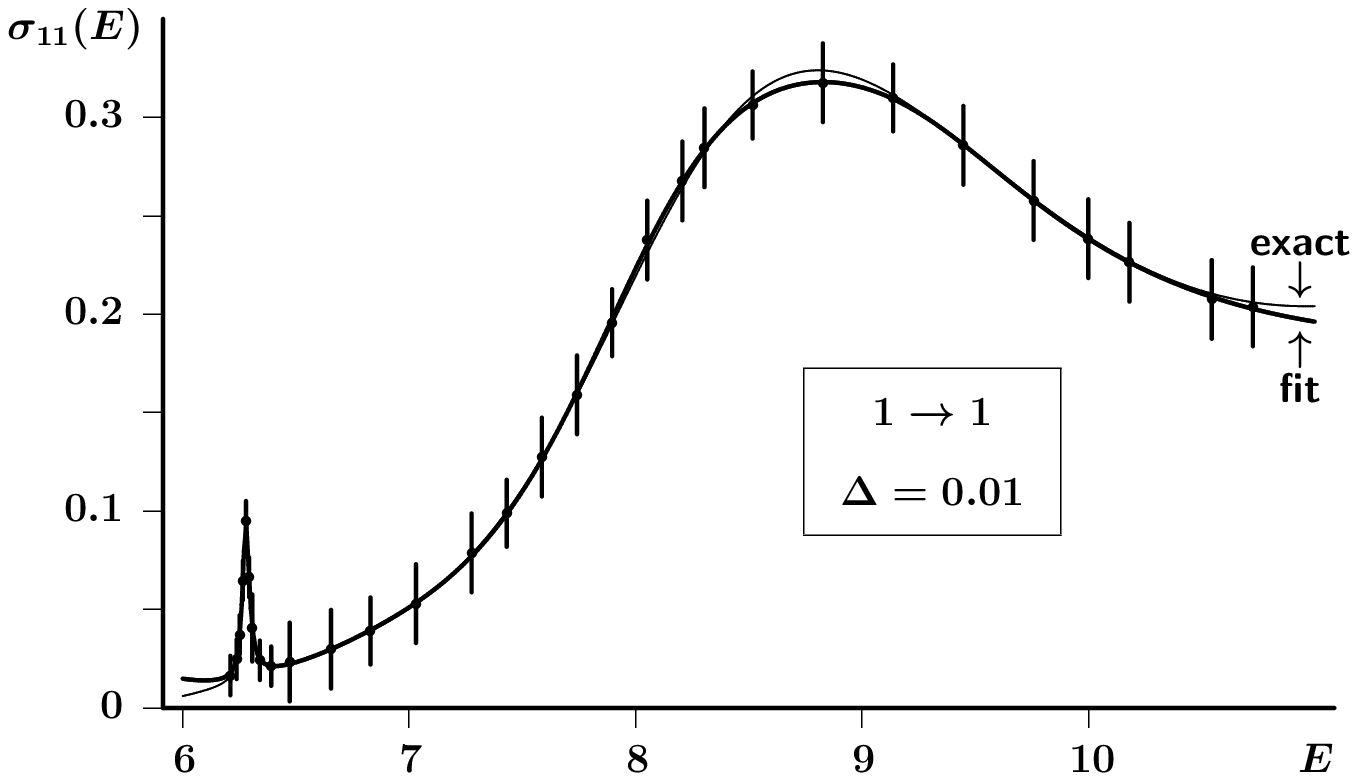}}
\caption{\sf
The data points for the elastic channel $(1\to1)$ together with the curves
showing the exact and fitted cross sections. The experimental noise for the
points has the normal distribution with the standard deviation $\Delta=0.01$.
}
\label{fig.11_0.01}
\end{figure}

\begin{figure}[ht!]
\centerline{\epsfig{file=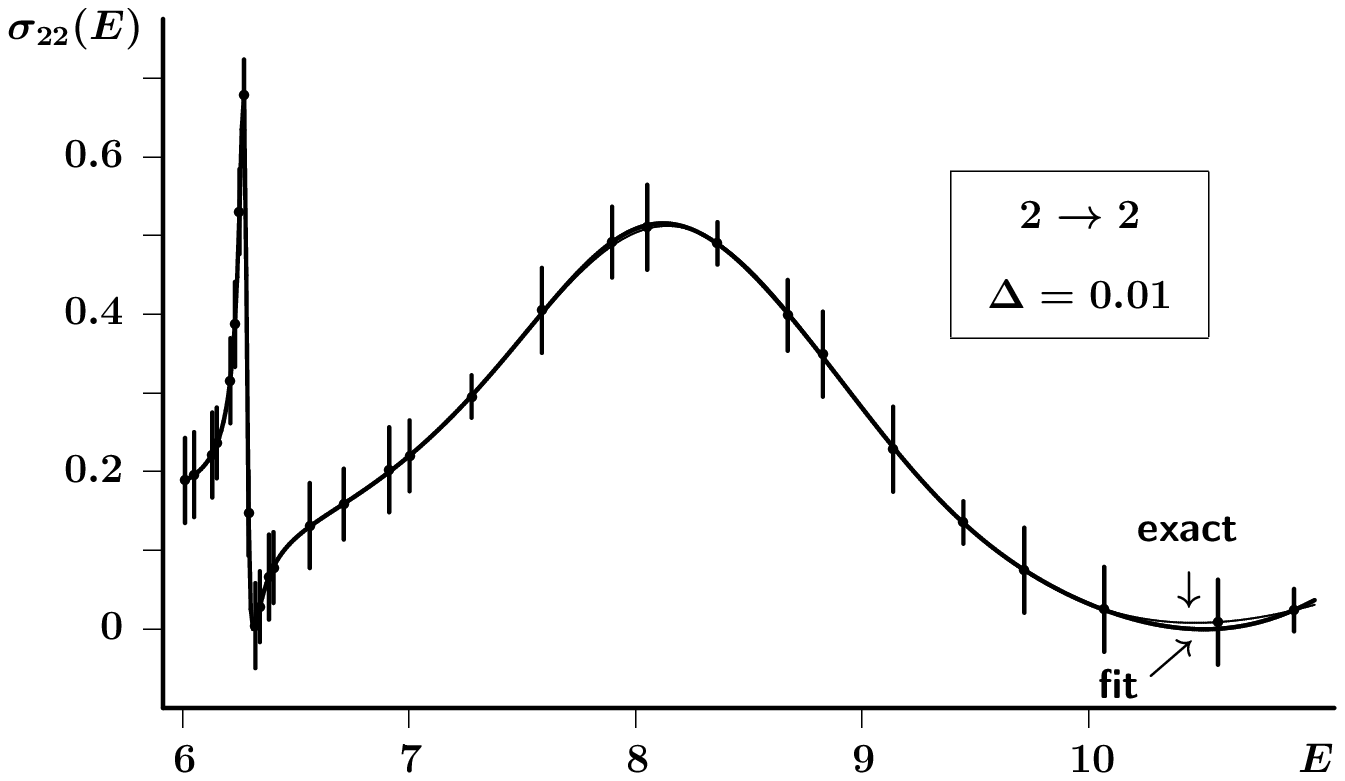}}
\caption{\sf
The data points for the elastic channel $(2\to2)$ together with the curves
showing the exact and fitted cross sections. The experimental noise for the
points has the normal distribution with the standard deviation $\Delta=0.01$.
}
\label{fig.22_0.01}
\end{figure}

\begin{figure}[ht!]
\centerline{\epsfig{file=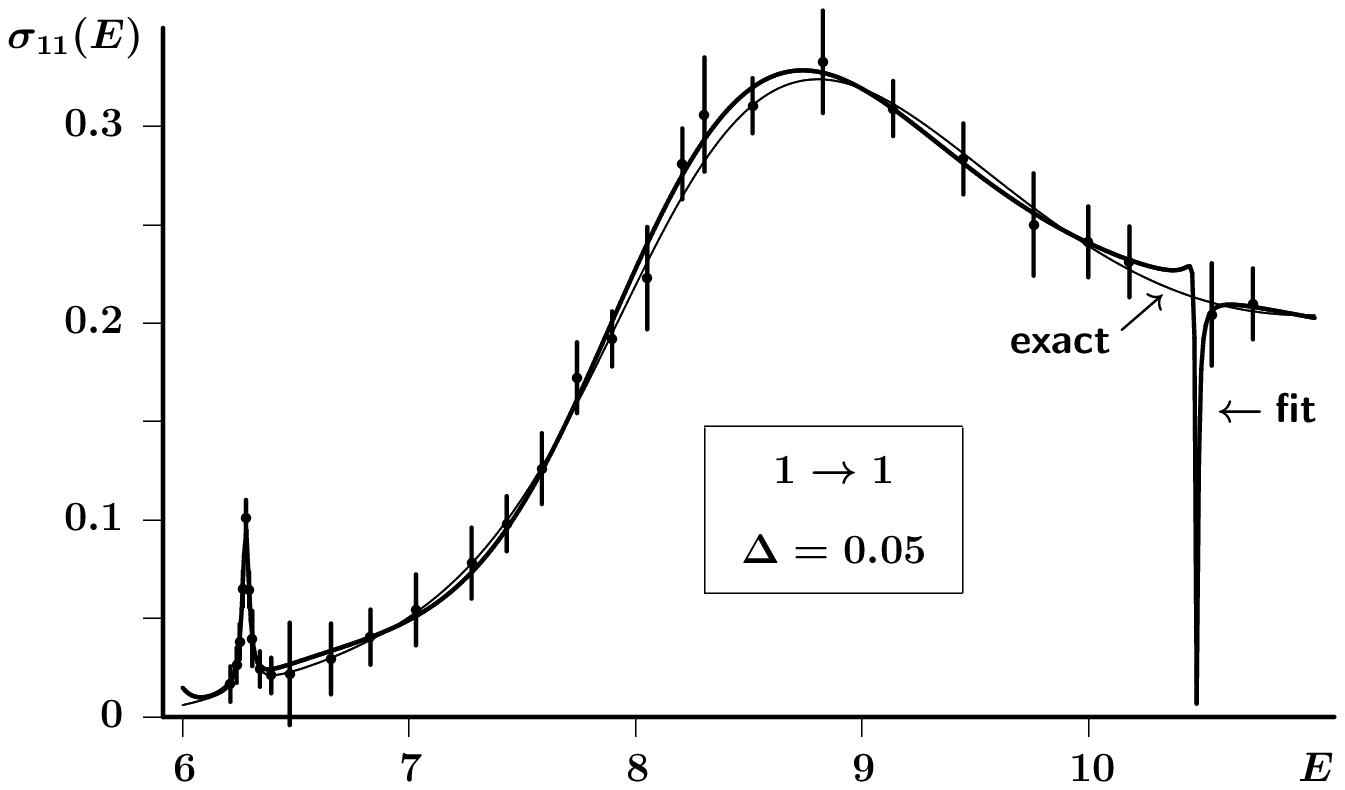}}
\caption{\sf
The data points for the elastic channel $(1\to1)$ together with the curves
showing the exact and fitted cross sections. The experimental noise for the
points has the normal distribution with the standard deviation $\Delta=0.05$.
}
\label{fig.11_0.05}
\end{figure}

\begin{figure}[ht!]
\centerline{\epsfig{file=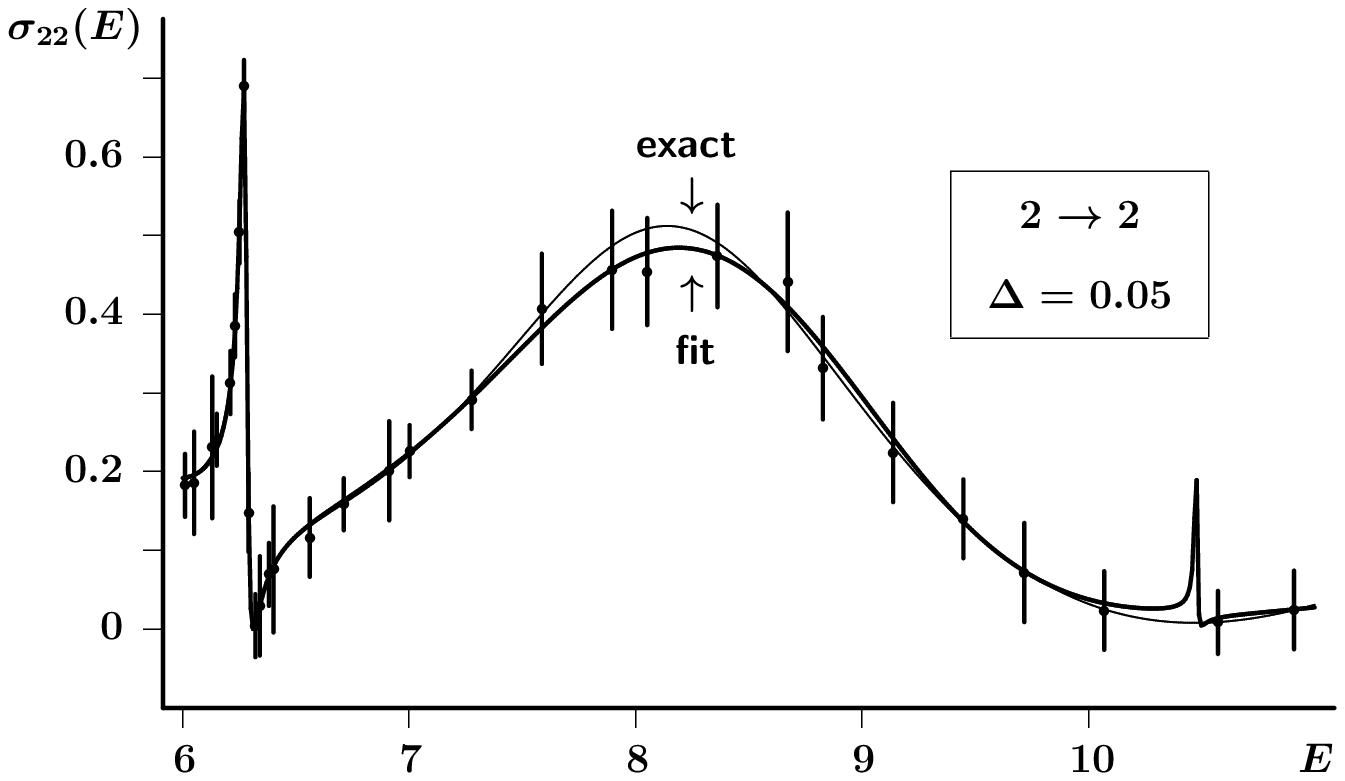}}
\caption{\sf
The data points for the elastic channel $(2\to2)$ together with the curves
showing the exact and fitted cross sections. The experimental noise for the
points has the normal distribution with the standard deviation $\Delta=0.05$.
}
\label{fig.22_0.05}
\end{figure}

\begin{figure}[ht!]
\centerline{\epsfig{file=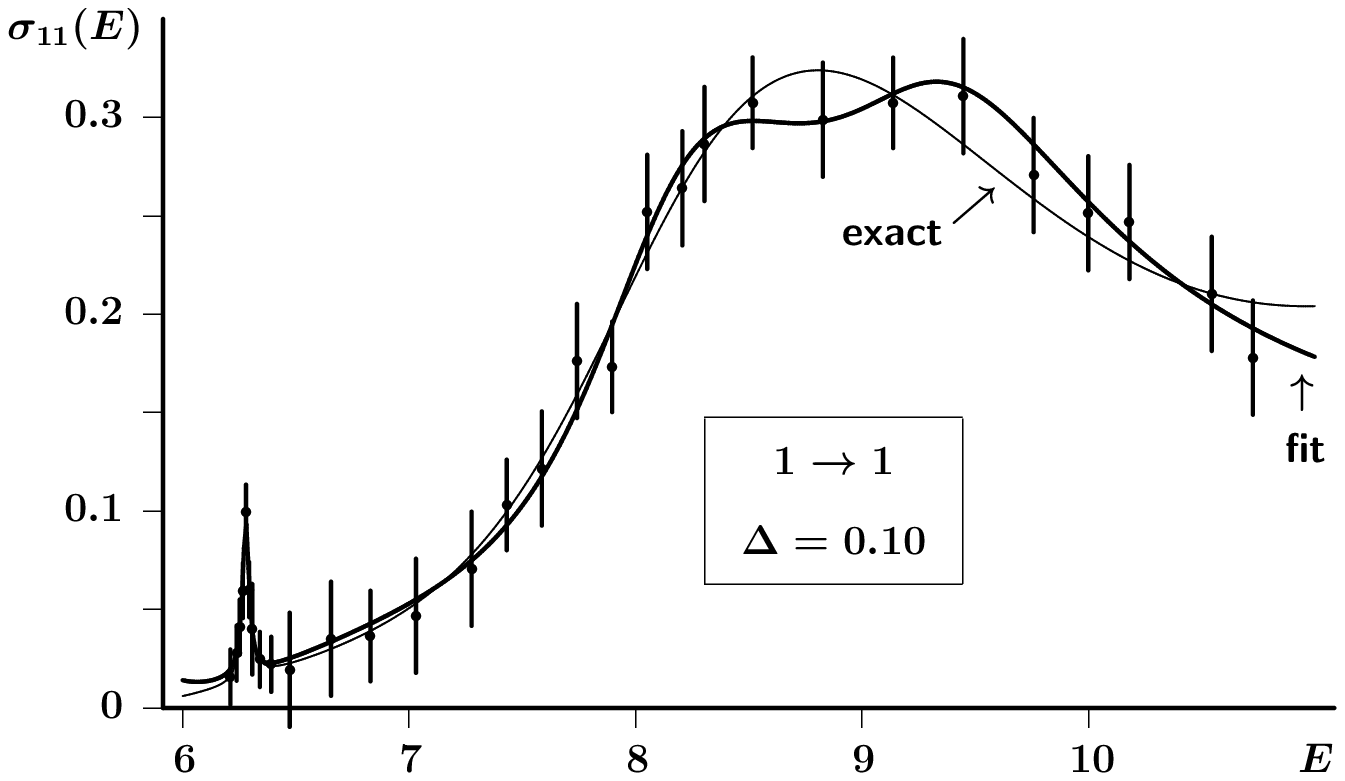}}
\caption{\sf
The data points for the elastic channel $(1\to1)$ together with the curves
showing the exact and fitted cross sections. The experimental noise for the
points has the normal distribution with the standard deviation $\Delta=0.10$.
}
\label{fig.11_0.10}
\end{figure}

\begin{figure}[ht!]
\centerline{\epsfig{file=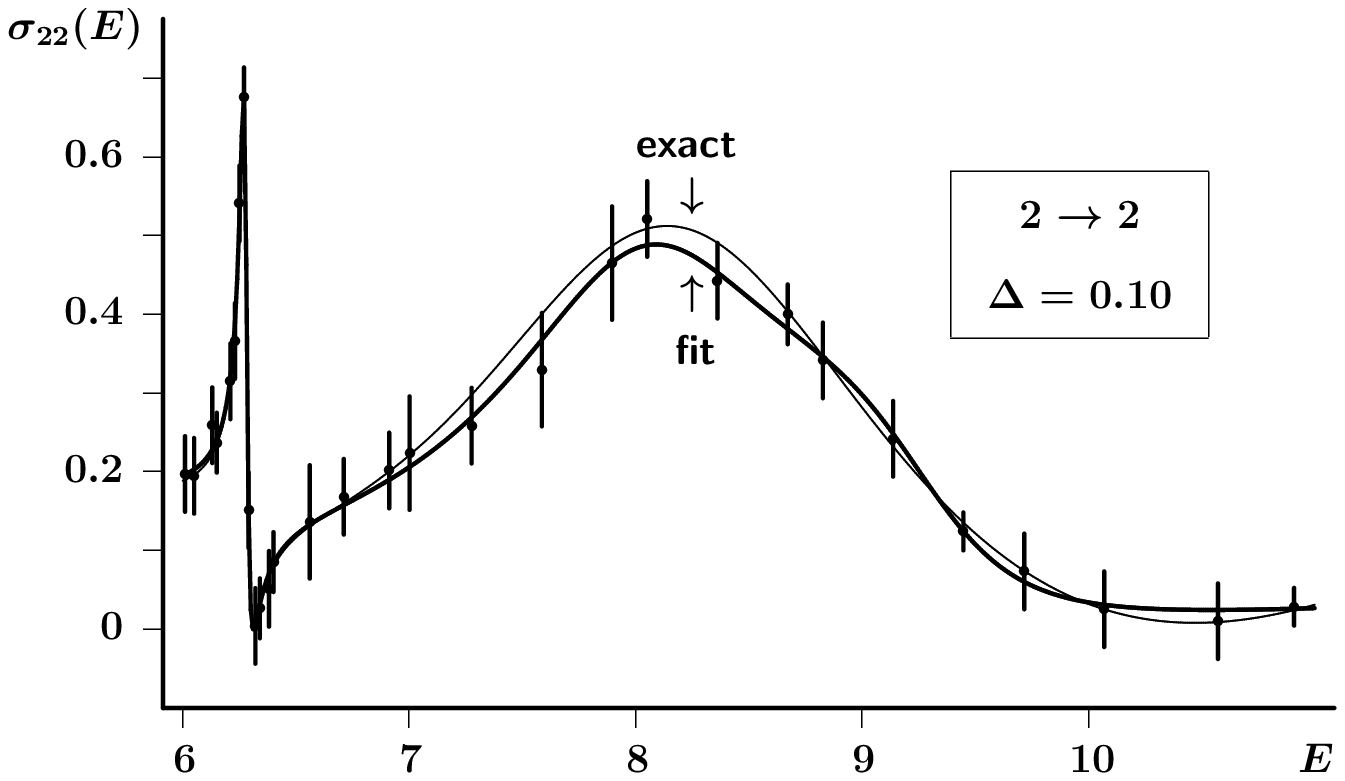}}
\caption{\sf
The data points for the elastic channel $(2\to2)$ together with the curves
showing the exact and fitted cross sections. The experimental noise for the
points has the normal distribution with the standard deviation $\Delta=0.10$.
}
\label{fig.22_0.10}
\end{figure}

Fig.~\ref{fig.11_0.01} shows the exact cross section $\sigma_{11}(E)$, the
artificial data points with $\Delta=0.01$, and the curve obtained by fitting
these points. The same information for the channel $(2\to2)$ is given in
Fig.~\ref{fig.22_0.01} (also for $\Delta=0.01$). Similarly,
Figures \ref{fig.11_0.05}, \ref{fig.22_0.05} and \ref{fig.11_0.10},
\ref{fig.22_0.10} show the corresponding exact and fitted cross sections as well
as the data points for stronger experimental noise, namely, for $\Delta=0.05$
and $\Delta=0.10$.\\

\begin{figure}[ht!]
\centerline{\epsfig{file=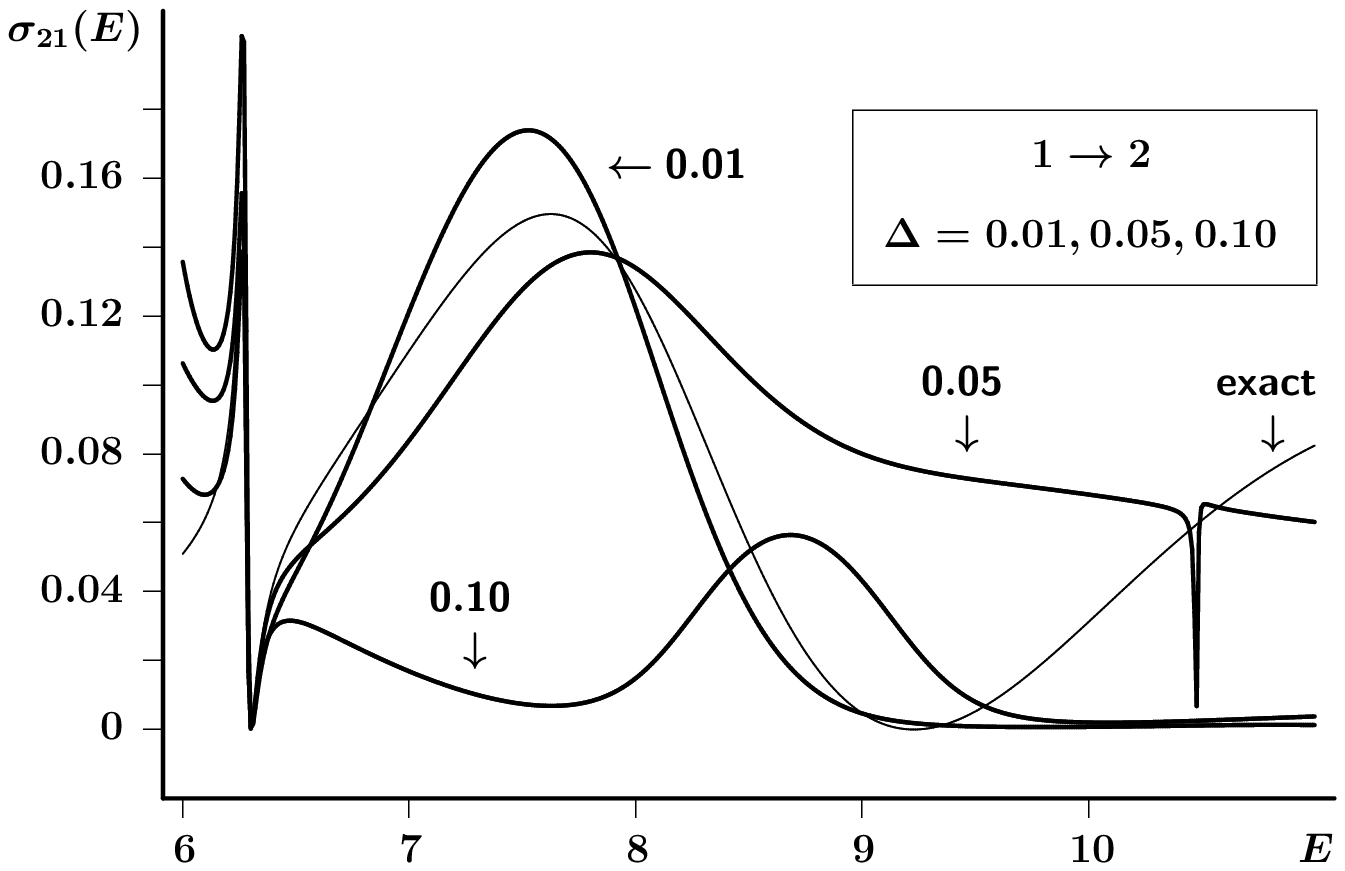}}
\caption{\sf
Exact  inelastic cross section (thin curve) for the channel $(1\to2)$ and the
approximate curves obtained after fitting the data in the elastic channels with
the experimental noise determined by the standard deviations $\Delta=0.01, 0.05,
0.10$.
}
\label{fig.21_all}
\end{figure}

In fitting the data points in the elastic channels $(1\to1)$ and
$(2\to2)$ all matrix elements of the Jost matrices are involved. As a result,
not only the diagonal but also the off-diagonal elements of the $S$-matrix
should be close to the correct values. This means that even without having any
data points in the inelastic channels, we should obtain the cross sections
$\sigma_{21}(E)$ and $\sigma_{12}(E)$ that are not far from the corresponding
exact curves. Fig.~\ref{fig.21_all} shows the exact cross section for the
inelastic channel $(1\to2)$ and the curves obtained for it with
$\Delta=0.01,\ 0.05,\ 0.10$. Of course, as one would expect, the greater the
accuracy of the experimental data, the more accurate is the prediction for the
cross section in the channel where no data are available.\\

After fitting the data, we looked for the roots of Eq.~(\ref{spectral}) on the
non-physical sheet of the Riemann surface of the energy. For this sheet, the
signs in Eq.~(\ref{chmom}) are chosen in such a way that both channel
momenta $k_1$ and $k_2$ have negative imaginary parts. The roots thus found
correspond to the resonance spectral points. They are listed in
Table~\ref{table.fitted_poles} for all three levels of the experimental noise.
Of course the more accurate the measurements, the closer to the exact
values are the extracted resonance parameters. Even with very
high experimental noise ($\Delta=0.10$) we are still able to extract at least
the first resonance with a reasonable accuracy.
\begin{table}[htbp]
\centering 
\begin{tabular}{| c || c || c | c | c | c |} 
\hline
Resonance & $\Delta$ & $E_{r}$ & $\Gamma$ & $\Gamma_{1}$ & $\Gamma_{2}$\\[1ex]
\hline\hline
   & exact & 6.278042552 & 0.036866729 & 0.006898807 & 0.029967922 \\
1  & 0.01  & 6.277997424 & 0.036731019 & 0.006721542 & 0.030009477 \\
   & 0.05  & 6.278563562 & 0.035568397 & 0.006497720 & 0.029070677 \\
   & 0.10  & 6.278669302 & 0.036236713 & 0.006638945 & 0.029597768 \\   [1ex]
\hline
  & exact & 8.038507867 & 2.563111275 & 0.617710684 & 1.945400591 \\
2 & 0.01  & 7.998939904 & 2.096675299 & 0.623726003 & 1.472949296 \\
  & 0.05  & 7.676616089 & 2.502856671 & 0.792088450 & 1.710768220 \\
  & 0.10  & 7.968634195 & 1.662113407 & 0.231505793 & 1.430607614 \\ [1ex]
\hline
  & exact & 8.861433400 & 7.883809114 & 1.949506410 & 5.934302704 \\
3 & 0.01  & 11.21325906 & 3.204531546 & 0.031776330 & 3.172755216  \\
  & 0.05  & 9.188805831 & 2.549030291 & 0.364986606 & 2.184043685  \\
  & 0.10  & 9.259323135 & 2.226793463 & 1.232709401 & 0.994084062 \\ [1ex]
\hline
\end{tabular}
\caption{\sf
The resonance parameters obtained from fitting the data with
different degrees of experimental noise.}
\label{table.fitted_poles}
\end{table}

\section{Conclusion}
The main idea of the proposed method has its roots in the so called
``effective-range expansion'' widely used in nuclear and atomic physics (see,
for example, Ref.~\cite{Kukulin}). Within this approach, a certain function of
the scattering phase-shift is expanded in the power series of the collision
energy, and the expansion coefficients are used as the adjustable parameters to
fit experimental data. Traditional effective-range expansion is very useful but
is limited to low energies. Moreover, it is difficult to apply it to
multi-channel processes.\\

We generalize this approach and thus remove all these limitations. First of
all, we expand a more fundamental quantity: the Jost matrix. When the
Jost-matrix expansion in the power series is obtained, one can easily derive the
corresponding expansion of the $S$-matrix, the phase-shift, or any other
quantity that is needed. Secondly, it is not necessary to do the expansion near
the point $E=0$. Actually, the expansion can be done around any complex value of
variable $E$. Concerning the multi-channel problems, for the Jost-matrices, this
does not pose any difficulty. Simply, the expansion coefficients become
matrices.\\

In order to expand a function in the Taylor series of $E$, one has to be
sure that this function is an analytic and single-valued function of $E$.
However, all the quantities describing the scattering processes (amplitude,
$S$-matrix, Jost matrix, etc) are multi-valued functions defined on a
complicated Riemann surface of the energy with the number of branch points equal
to the number of channels. In order to circumvent this difficulty, we use
earlier derived general semi-analytic expression for a multi-channel
Jost matrix, where all the factors responsible for its ``bad behaviour'', are
given explicitly. The remaining unknown functions are more simple, more smooth,
and (more importantly) are analytic and single-valued
functions defined on a simple energy plane. We only do the expansion for
these functions.\\

In our earlier publication \cite{ourFit}, we demonstrated how the proposed
method works for the reactions involving neutral particles. In the present
paper, we consider a more general case when the potential has both a
short-range part and a Coulomb tail. Similarly to the simple effective-range
theory, presence of the Coulomb potential makes the explicit coefficients
in the semi-analytic expression for the Jost matrix more complicated.
However, the remaining functions of $E$ are still smooth and can be
approximated by just a few terms of the Taylor series. We have demonstrated this
using a two-channel model.\\

We have shown that even with rather inaccurate experimental data ($\Delta=0.10$
in our model) the resonances can still be found. For the narrow resonance, we
managed to reproduce not only the energy and total width, but the partial
widths as well, when the data points had rather big deviations from the exact
values. For a wide resonance (the resonance number 2) with such a high
experimental noise, we still obtained  reasonable parameters. And even the
energy of the extremely wide resonance (number 3) was obtained not far from the
exact value. This shows that the proposed method is accurate and stable.\\

One of the advantages of the proposed method is that the fitting procedure
involves all matrix elements of the Jost matrix and therefore all elements of
the $S$-matrix even if the data are available just in one channel. The
resulting $S$-matrix should therefore be correct in all channels. This means
that by fitting accurately-measured data in one or two channels, we could in
principle obtain a reasonable estimate for the cross section in the other
channels where the measurements are difficult or impossible.\\

It should be noted that the proposed method is non-relativistic and therefore
cannot be directly used in high-energy physics. There are however a wide
range of problems in atomic and low-energy nuclear physics, where it could find
applications. In principle, one can try the same parametrization for high
energies as well, if the relativistic relation between the energy and momentum,
$E=\sqrt{\hbar^2k^2c^2+\mu^2c^4}$, is used in all the formulae. Such
``intuitive'' inclusion of relativistic
kinematics into non-relativistic operators is very common for various
parametrizations of scattering data in particle physics. In our case, however,
this would mean that the mathematical rigor and substantiation are lost.


\end{document}